# Anisotropic superconductivity in epitaxial $MgB_2$ films

M.H. Jung [a,*], M. Jaime [a], A.H. Lacerda [a], G.S. Boebinger [a], W.N. Kang [b], H.J. Kim [b], E.M. Choi [b], S.I. Lee [b]

[a]*National High Magnetic Field Laboratory, Los Alamos National Laboratory, MS E536 Los Alamos, NM 87545, USA*
[b]*National Creative Research Initiative Center for Superconductivity and Department of Physics, Pohang University of Science and Technology, Pohang 790-784, South Korea*

## Abstract

High-quality epitaxial $MgB_2$ thin films prepared by pulsed laser deposition with $T_c$ = 39 K offer the opportunity to study the anisotropy and robustness of the superconducting state in magnetic fields. We measure the in-plane electrical resistivity of the films in magnetic fields to 60T and estimate the superconducting upper critical field $H_\rho{}^c(0) \approx 24 \pm 3$ T for field oriented along the *c*-axis, and $H_\rho{}^{ab}(0) \approx 30 \pm 2$ T for field in the plane of the film. We find the zero-temperature coherence lengths $x_c(0) \approx 30$ Å and $x_{ab}(0) \approx 37$ Å to be shorter than the calculated electronic mean free path $l \approx 100 \pm 50$ Å, which places our films in the clean limit. The observation of such large upper critical fields from clean limit samples, coupled with the relatively small anisotropy, provides strong evidence of the viability of $MgB_2$ as a technologically important superconductor.

*Corresponding author. Fax: +1-505-665-4311
E-mail address: mhjung@lanl.gov (M.H. Jung).

## 1. Introduction

The recent discovery of superconductivity in $MgB_2$ has brought a renewal of interest in intermetallic superconductivity because of the high transition temperature $T_c$ = 39 K [1]. The only other superconductors at these temperatures are doped fullerenes [2] and high-$T_c$ cuprates [3]. The extreme sensitivity to oxidation of the fullerenes severely limits their technological applications. On the other hand, the small superconducting coherence lengths and the strongly two-dimensional character in the cuprates result in unwanted vortex motion dissipation, that can only be avoided with complicated and expensive texture processes designed to pin the highly mobile vortices in anisotropic materials. Polycrystals and wires of $MgB_2$, a new candidate for long-awaited high-temperature superconductivity applications, show disappointingly small critical magnetic fields $H_{c2}$ = 12 – 16 T [4,5], which is smaller than the $Nb_3Sn$ alloy ($T_c$ = 18 K, $H_{c2}$ = 30 T) currently used in high field applications [6]. In the present study, we report the first transport measurements that show critical fields in high-quality epitaxial $MgB_2$ thin films to be comparable to those in $Nb_3Sn$. Moreover, we find evidence in our data indicating that these films are in the clean limit and have relatively small anisotropy, both facts which suggest technological applications for $MgB_2$ upon the introduction of artificial defects and/or texture to maximize vortex pinning. With a transition temperature as high as that of $MgB_2$, inexpensive cryogen-free refrigeration is sufficient to access the superconducting phase. With only slight anisotropy in the superconducting state, efficient pinning of vortex motion will likely be possible, offering the hope of dissipationless superconducting transport at magnetic fields comparable to the upper critical fields estimated from our transport measurements. This makes $MgB_2$ a viable challenger for $Nb_3Sn$, the most technologically important superconductor for high-field superconducting solenoids.

## 2. Experimental

We have recently succeeded in growing high-quality epitaxial $MgB_2$ thin films deposited on oriented (1102) $Al_2O_3$ substrates using ex-situ pulsed laser deposition [7]. The $MgB_2$ epitaxial films examined in this study have a thickness of 4000 Å and show a superconducting transition temperature $T_c$ = 39 K, measured by resistivity (see Fig. 1) and magnetization versus temperature. X-ray diffraction patterns (inset of Fig. 1) taken before and after film deposition indicate that $MgB_2$ films are epitaxially aligned along the *c* axis. The hexagonal lattice constants are $a_o$ = 3.10 and $c_o$ = 3.52 Å, close to those previously reported for sintered polycrystals [1]. A detailed study of our $MgB_2$ thin films includes scanning electron microscopy (SEM) images that show a continuous film with no discontinuity at the interface between the substrate and the $MgB_2$ films [7].

The in-plane resistivity of our films shows small residual resistivity $r_{ab}$(40K) ≈ 5.1 μΩcm and metallic behavior in the normal state (Fig. 1) with a dominant square of temperature ($T^2$) dependence and a residual resistivity ratio *RRR* = 2.5. A larger *RRR* has been observed in sintered polycrystalline $MgB_2$ samples [8]; however, the residual resistivity $r_{ab}$(40K) of our $MgB_2$ films is similar to the smallest values reported for sintered polycrystals. The sharp (0.7 K) superconducting transition at $T_c$ = 39 K in our films is comparable to that found in both polycrystalline samples [4,8] and $MgB_2$ wires [5,9]. The strength of the $T^2$ term of the electrical resistivity in our films indicates strong electron-electron (*e-e*) scattering and hence a long quasiparticle mean free path: the data of Fig. 1 can be described with an expression of the form $r_{ab}(T) = r_o + a\,T^2 + b\,T^5$. From a least squares fit we obtain $r_o = 4.89 \times 10^{-6}$ Ωcm, $a = 5.04 \times 10^{-11}$ ΩcmK$^{-2}$, and $b = 3.8 \times 10^{-18}$ ΩcmK$^{-5}$. The $T^5$ term, usually attributed to electron-phonon scattering in the weak coupling limit, only accounts for 5% of the total resistivity at 150 K in our films and, thus, is negligible.

Evidence for *e-e* scattering in $MgB_2$ comes from the empirical Kadowaki-Woods (KW) relationship [10] between the $T^2$-dependent coefficient $a$ and the Sommerfeld coefficient $g$ in the specific heat. Using the electrical resistivity data (Fig. 1) and available specific heat data [11,12], we find the ratio $a/g^2 = (2.5\pm1.5) \times 10^{-5}$ μΩcm (mole K/mJ)$^2$, quite close to the KW value $10^{-5}$ μΩcm (mole K/mJ)$^2$, which indicates a dominant *e-e* scattering and long quasiparticle mean free path.

Magnetoresistance measurements in a superconducting 20 T magnet and in a pulsed 60 T magnet were performed using the standard four-probe *AC* method with the current applied in the plane of the films. The magnetic field was applied perpendicular and along the film plane (*H*//*c* and *H*//*ab* configuration, respectively). In both configurations we find very small magnetoresistance in the normal state (< 2% at 18 T), consistent with reports from sintered polycrystalline samples [8], and in distinct contrast with $MgB_2$ wires for which the magnetoresistance is roughly 200 times larger [5]. The magnetoresistance measured at 45 K (Fig. 2) is found to increase monotonically up to 8% with an applied magnetic field to 60 T for *H*//*c* and up to 13% for *H*//*ab*. The observed anisotropy indicates smaller electronic effective mass in the plane of the film, that is $m_{ab} < m_c$. The normal-state magnetoresistance curves can be fitted to $\Delta R(H) = R(H) - R(0) = a\, R(0)\, H^b$ with $b = 2$, which is characteristic of simple metallic behavior, very well up to 30 T. Above 30 T, the data fall below the quadratic fit, which we attribute to the onset of the quantum regime, $w_c t > 1$, (where $w_c$ is the cyclotron frequency and $t$ is the scattering time) in which quantum oscillations are likely to be observed in single crystals of this material. This evidence of the quantum regime, together with the magnitude of the $T^2$ term in the resistivity, indicate that the $MgB_2$ films examined in the present study are of high quality and contain few defects or impurities.

The temperature dependence of the resistivity in various magnetic fields to 18 T is displayed in Fig. 3A and 3B, where the data were normalized to the resistance at 40 K. As expected, the superconducting phase is suppressed to lower temperatures with increasing magnetic fields. From each curve, the onset of the superconducting transition is defined as the temperature at which the resistivity first deviates from the normal-state value (indicated by arrows). We also characterize the transition temperature using the intersection of two lines, one temperature-independent line corresponding to the normal-state resistivity and a second line drawn tangent to the data at the midpoint of the superconducting transition.

The magnetic field dependence of the resistivity to 40 T at constant temperature down to 0.5 K is shown in Fig. 3C and 3D, where the data were normalized to the resistance in a field of 30 T. There is a distinct suppression of the superconducting phase for *H*//*c* and *H*//*ab* to lower magnetic fields with increasing temperature, consistent with the data taken at constant field. The data in Fig. 3 show that magnetic fields in excess of 27 T are required to completely suppress the superconducting state in $MgB_2$ at low temperatures when they are applied along the *c*-axis (*H*//*c*), while fields exceeding 32 T are required when applied in the film plane ($H//ab$). Note also the different transition width for $H//ab$, which is narrower than that for $H//c$. These observations are clear evidence for anisotropic superconductivity in $MgB_2$. From the data in Fig. 3C and 3D, we estimate the 'resistive' upper critical field from each magnetoresistance curve, obtaining the onset upper critical field $H_{\rho\ onset}$(T), defined as the field at which the resistivity first deviates from the normal-state value (indicated by arrows), and $H_{\rho\ tangent}$(T) defined as the intersection of a line tangent to the data at the midpoint of the transition with a linear extrapolation of the normal-state resistivity. While we recognize that dissipation due to vortex motion can cloud efforts to determine the true upper critical magnetic field from transport measurements (hence we use the term 'resistive' upper critical field and use the $H_\rho$ (T) notation), we note that transport measurements provide a *lower bound* for the upper critical magnetic field [3] and that this lower bound is of surprisingly large magnitude.

Figures 4A and 4B show both $H_{\rho\ onset}$(T) and $H_{\rho\ tangent}$(T) when the magnetic field is in the film plane and perpendicular to the film plane. There is a marked upward curvature at temperatures above 34 and 29 K for $H//c$ and $H//ab$, respectively. This upward curvature, observed in polycrystalline $MgB_2$ samples [4,5,8,9], in other boride superconductors such as $LuNi_2B_2C$ and $YNi_2B_2C$ [13], as well as in some high-$T_c$ superconductors [14], makes it unreliable to estimate the zero-temperature critical field $H_{c2}(0)$ from the data taken near $T_c$ in low magnetic fields. We therefore employed pulsed magnetic fields up to 60T to measure the critical fields down to the zero-temperature limit. From our two criteria discussed above we estimate the resistive upper critical fields $H_\rho(T) = [H_{\rho\ onset}(T) + H_{\rho\ tangent}(T)]/2$ to be $H_\rho^c(0) \approx 24 \pm 3$ T and $H_\rho^{ab}(0) \approx 30 \pm 2$ T as the temperature approaches zero, where the bulk of the uncertainty results from the two different characterizations of the transition that we employ.

### 3. Results and Discussion

In order to see quantitatively the anisotropy in $MgB_2$, we plot the ratios $H_{\rho\ onset}{}^{ab}(T)/H_{\rho\ onset}{}^{c}(T)$ and $H_{\rho\ tangent}{}^{ab}(T)/H_{\rho\ tangent}{}^{c}(T)$, which are found to be temperature independent in Fig. 4C. Note that the observed anisotropy does not change much with the definition used for the resistive critical field. The Ginzburg-Landau theory [3] for anisotropic superconductors gives for magnetic field applied along the *c* axis $H_{c2}{}^{c}$ = $F_o/2px_{ab}{}^2$ and $x_{ab}{}^2 = h^2/2m_{ab}|a(T)|$, where $F_o$ is the flux quantum, $x_{ab}$ is the in-plane coherence length, $m_{ab}$ is the in-plane effective mass of the quasiparticles, and $a(T)$ is the Ginzburg-Landau parameter. For magnetic field applied in the *ab* plane, $H_{c2}{}^{ab}$ = $F_o/2px_{ab}x_c$ and $x_c{}^2 = h^2/2m_c|a(T)|$. Using these equations and our values for the upper critical fields we find $H_\rho{}^{ab}/ H_\rho{}^{c} = x_{ab}/x_c = (m_c/m_{ab})^{1/2} \approx 1.25 \pm 0.05$ at all temperatures below 32 K, Fig. 4C. This is a relatively small anisotropy given the layered structure of $MgB_2$, which bodes well for future prospects to efficiently pin the vortices for high temperature superconducting applications of $MgB_2$. Similar values for this anisotropy result from *AC* magnetic susceptibility measurements on micro-crystals of $MgB_2$ [15].

The values of $H_\rho{}^{c}$ (0) give zero-temperature coherence lengths of $x_c(0)$ = 30 Å and $x_{ab}(0)$ = 37 Å. These values can be used to determine that our $MgB_2$ films are in the clean limit, where the superconducting coherence length is smaller than the average distance between scattering of carriers. The electronic mean free path (*l*) in the normal state is estimated to be $l = 100 \pm 50$ Å in our films from the measured normal-state resistivity of 5 μΩcm (Fig. 1) and the reported Hall coefficient [16], using a free electron mass for the quasiparticles and a Fermi velocity $\nu_F = 4.8 \times 10^7$ cm/s from band calculations [17]. The data from our high-magnetic-field experiments place our $MgB_2$ epitaxial films within the clean limit since $l \geq x$. This also raises the likelihood of technological applications for $MgB_2$, since the purposeful introduction of artificial defects and/or texture to maximize vortex pinning will likely shift the dissipation-less transport regime in $MgB_2$ nearly up to (or even beyond) the values of the upper critical field estimated from our resistivity measurements.

In conclusion, we have prepared high-quality epitaxial $MgB_2$ thin films and characterized them by x-ray diffraction, magnetization, and electrical resistivity. The films are well aligned along the *c*-axis, have low resistivity, and exhibit a sharp superconducting transition at 39 K. We have measured the magnetoresistance to 60 T in both *H*//*c* and *H*//*ab* configurations and found the magnetoresistance proportional to $H^2$ up to the onset of the quantum limit at $H \approx 30$ T. We observe large upper critical fields $H_\rho^c(0) \approx 24 \pm 3$ T and $H_\rho^{ab}(0) \approx 30 \pm 2$ T, and a relatively small temperature independent anisotropy in the superconducting state. Estimates for the superconducting coherence length and electronic mean free path in the normal state indicate that the films are in the clean limit. These measurements provide strong evidence of promising technological applications for high-temperature and high-magnetic-field applications of the superconductivity in $MgB_2$.

**Acknowledgements**

Work at the National High Magnetic Field Laboratory was performed under the auspices of the National Science Foundation, the State of Florida and the U. S. Department of Energy. This work is supported by the Ministry of Science and Technology of Korea through the Creative Research Initiative Program. We thank K. H. Ahn and M. Maley for useful discussions. MHJ acknowledges partial support from LANSCE - LANL.

**Figure captions**

Fig. 1. Temperature dependence of electrical resistivity of the $MgB_2$ epitaxial thin film showing a sharp superconducting transition at 39 K. The solid line is a fit with the expression $r_{ab}(T) = r_o + a T^2 + b T^5$. The inset presents a x-ray diffraction pattern measured at room temperature, showing (001) peak of the epitaxial $MgB_2$ thin film as well as the peaks indexed as (1102) from the $Al_2O_3$ substrate. We also plot the x-ray spectrum for substrate alone, which does not show the peak attributed to $MgB_2$.

Fig. 2. Normal-state magnetoresistance $\Delta R/R(0) = [R(H)-R(0)]/R(0)$ measured at 45K for magnetic field applied in and perpendicular to the film plane.

Fig. 3. Resistance versus temperature for *H*//*ab* (A) and *H*//*c* (B) in various magnetic fields 0, 1, 2.5, 5, 10, 15, and 18 T from right to left. The data were normalized to the resistance at 40 K. Magnetoresistance for *H*//*ab* (C) at temperatures 1.5, 8, 12, and 15 K from right to left and for *H*//*c* (D) at temperatures 0.5, 6, 10, 15, 20, and 25 K from right to left. The data were normalized to the resistance at 30 T. The arrows indicate the onset temperatures (A, B) and fields (C, D) of superconducting transition.

Fig. 4. Upper critical field from onset of resistive transition $H_{\rho\ \mathrm{onset}}$ ( ■ and □ ) and midpoint tangent extrapolation $H_{\rho\ \mathrm{tangent}}$ ( ● and ○ ) obtained for *H*//*c* (A) and *H* //*ab* (B): Open symbols from *R*(*H*) measured at constant temperature and solid symbols from *R*(*T*) measured at constant field. (C) The ratio of the upper critical fields $H_{\rho\ \mathrm{onset}}^{ab}/H_{\rho\ \mathrm{onset}}^{c}$ ( □ ) and $H_{\rho\ \mathrm{tangent}}^{ab}/H_{\rho\ \mathrm{tangent}}^{c}$ ( ● ) with the error bars as a function of temperature.

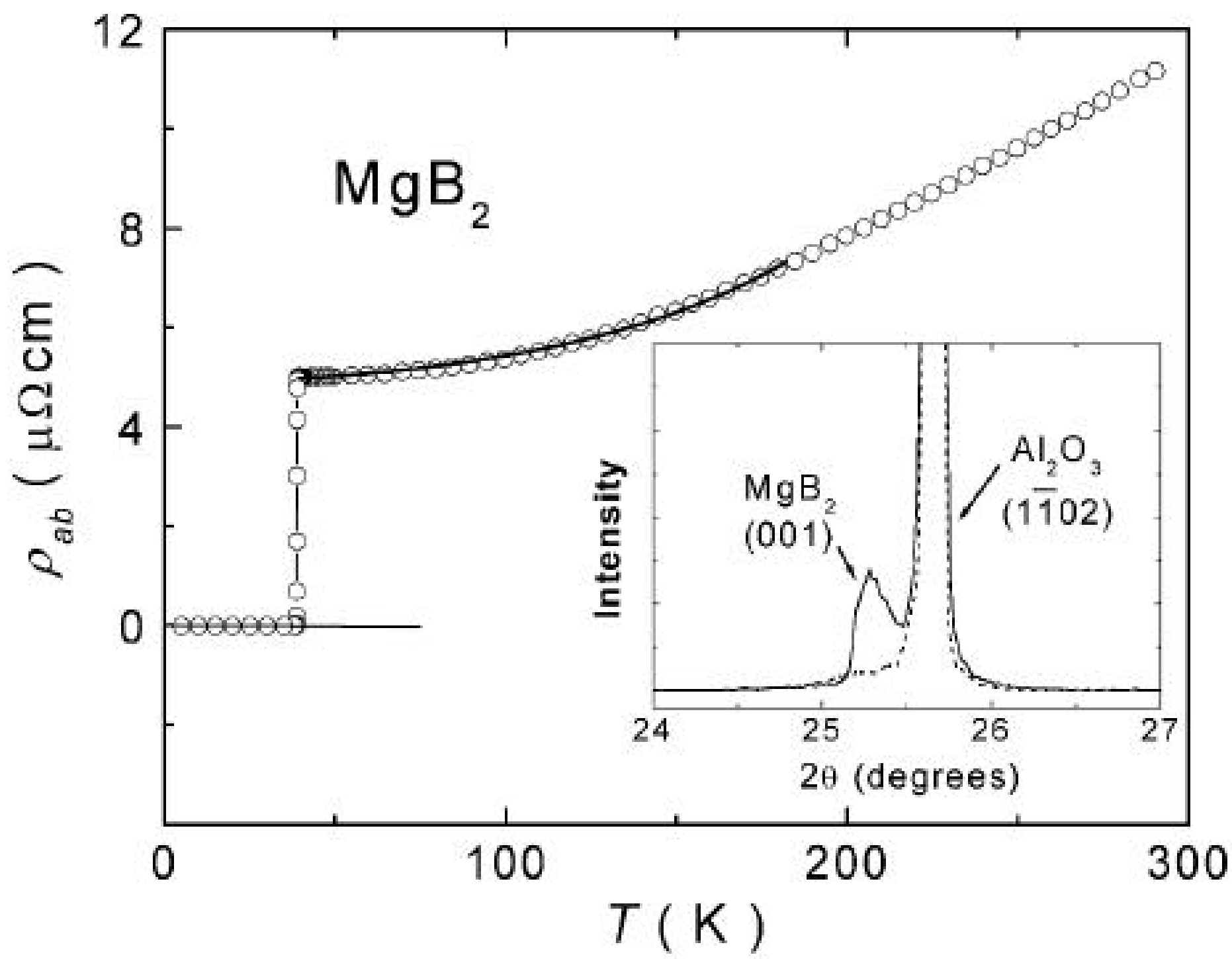
MgB$_2$
$\rho_{ab}$ ( μΩcm )
T ( K )
Intensity
MgB$_2$ (001)
Al$_2$O$_3$ (1$\bar{1}$02)
2θ (degrees)

Fig 1

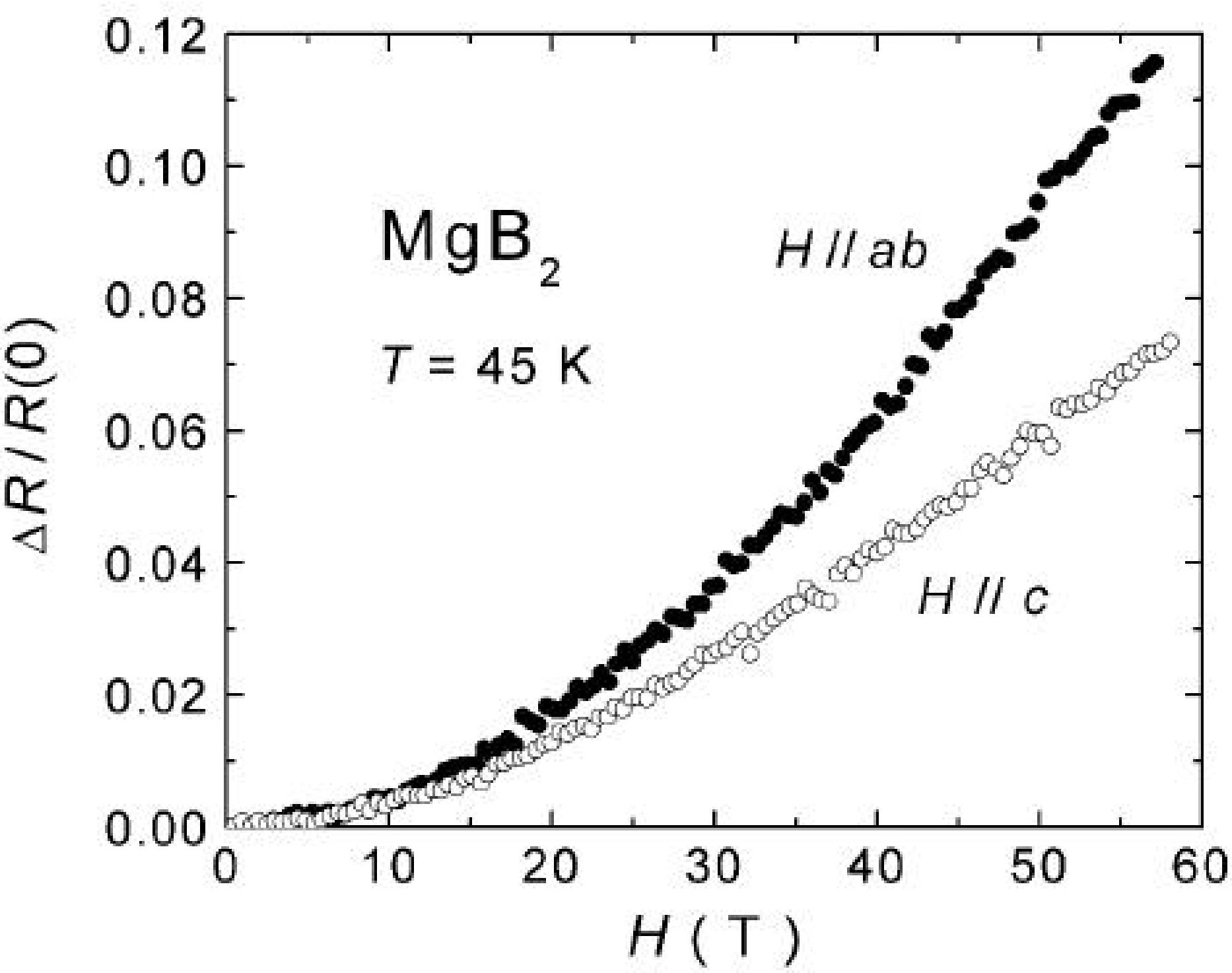
MgB$_2$
T = 45 K
H // ab
H // c
ΔR / R(0)
H ( T )

Fig 2

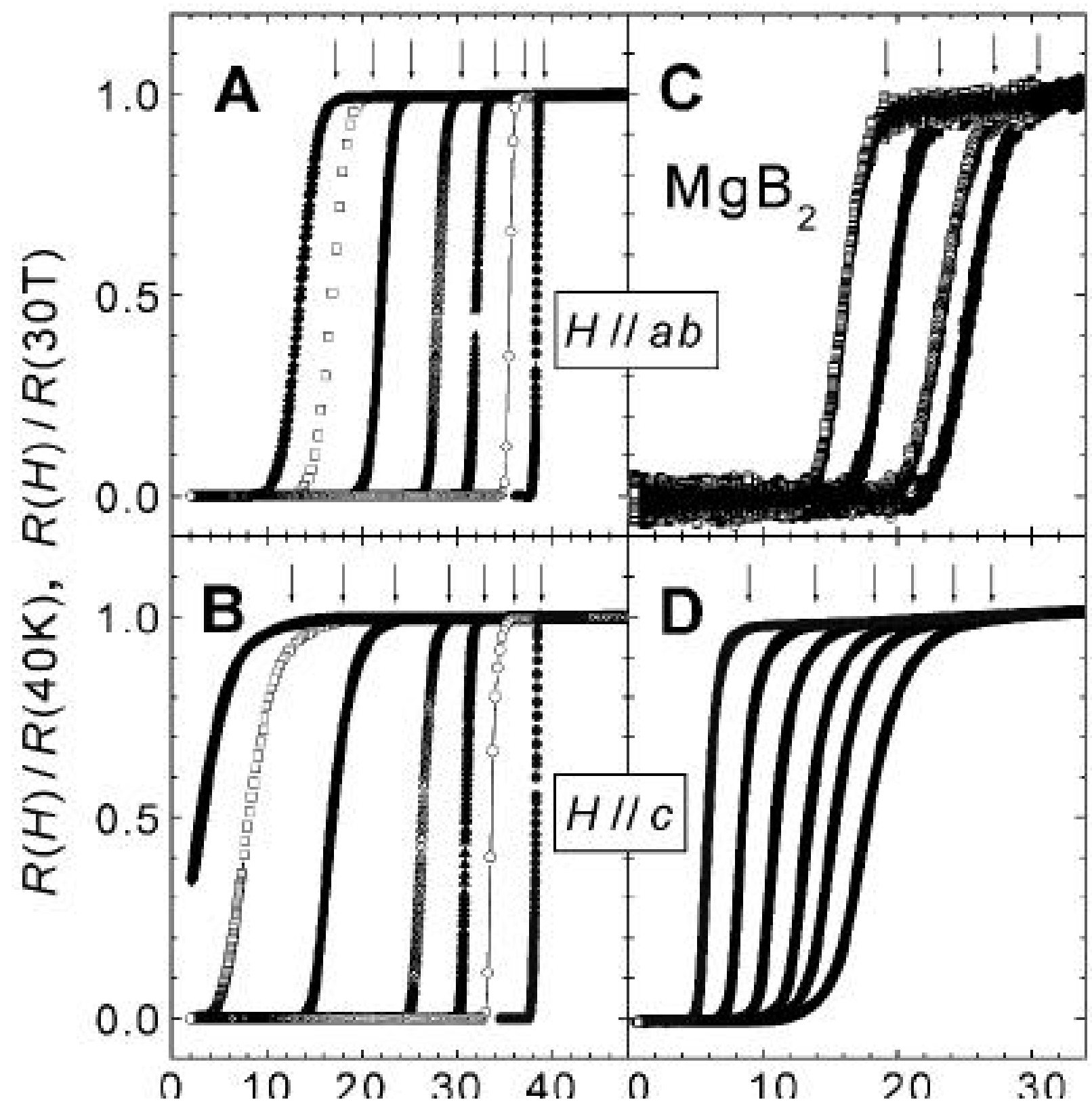

A
B
C
D
MgB$_2$
H // ab
H // c
R(H) / R(40K), R(H) / R(30T)
1.0
0.5
0.0
0
10
20
30
40

Fig 3

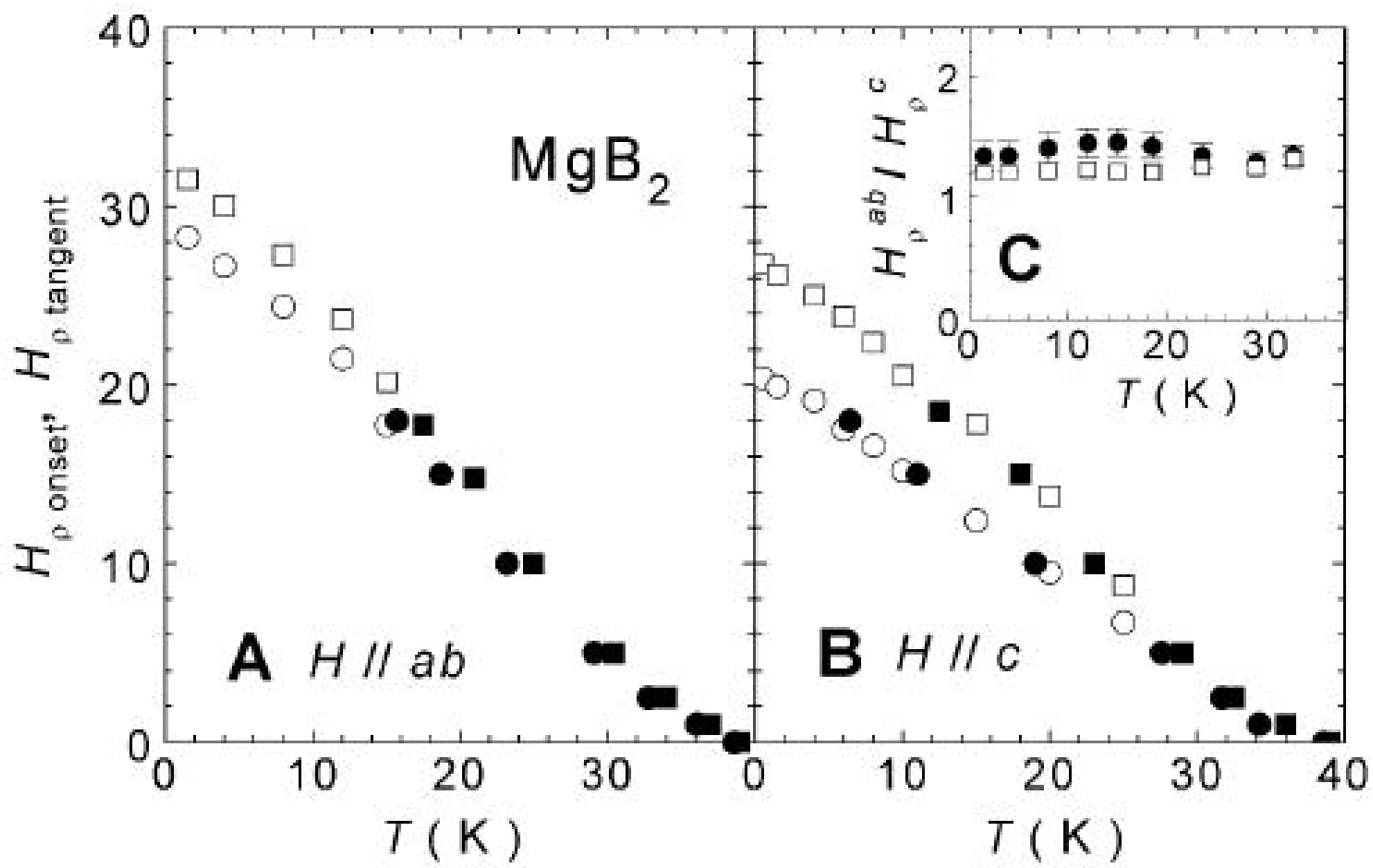

MgB$_2$
A H // ab
B H // c
C
$H_{\rho\ onset}$, $H_{\rho\ tangent}$
$H_\rho^{ab}$ / $H_\rho^c$
T ( K )
0
10
20
30
40

Fig 4